\documentclass[twoside,twocolumn,english,aps,amsmath,amssymb,reprint,pra,superscriptaddress]{revtex4-1}
\usepackage[T1]{fontenc}
\usepackage[latin9]{inputenc}
\usepackage{color}
\usepackage{babel}
\usepackage{varioref}
\usepackage{amsmath}
\usepackage{amssymb}
\usepackage{graphicx}

\makeatletter
\usepackage{babel}
\usepackage{babel}

\makeatother

\begin{document}
\title{Quantum aggregation with temporal delay}
\author{Nicolo Lo Piparo}
\email{nicopale@gmail.com}

\affiliation{Okinawa Institute of Science and Technology Graduate University, 1919-1
Tancha, Onna-son, Okinawa, 904-0495, Japan.}
\author{William J. Munro}
\affiliation{Okinawa Institute of Science and Technology Graduate University, 1919-1
Tancha, Onna-son, Okinawa, 904-0495, Japan.}
\affiliation{National Institute of Informatics, 2-1-2 Hitotsubashi, Chiyoda-ku,
Tokyo 101-8430, Japan.}
\author{Kae Nemoto}
\affiliation{Okinawa Institute of Science and Technology Graduate University, 1919-1
Tancha, Onna-son, Okinawa, 904-0495, Japan.}
\affiliation{National Institute of Informatics, 2-1-2 Hitotsubashi, Chiyoda-ku,
Tokyo 101-8430, Japan.}
\begin{abstract}
Advanced quantum networking systems rely on efficient quantum error
correction codes for their optimal realization. \textcolor{black}{The
rate at which the encoded information is transmitted is a fundamental
limit that affects the performance of such systems. Quantum aggregation
allows one to increase the transmission rate by adding multiple paths
connecting two distant users. Aggregating channels of different paths
allows more users to simultaneously exchange the encoded information.}
Recent work has shown that quantum aggregation can also reduce the
number of physical resources of an error correction code when it is
combined with the quantum multiplexing technique. However, the different
channel lengths across the various paths means some of the encoded
quantum information will arrive earlier than others and it must be
stored in quantum memories. The information stored will then deteriorate
due to decoherence processes leading to detrimental effects for the
fidelity of the final quantum state. Here, we explore the effects
of a depolarization channel that occurs for the quantum Reed-Solomon
code when quantum aggregation involving different channel lengths
is used. We determine the best distribution of resources among the
various channels connecting two remote users. Further we estimate
the coherence time required to achieve a certain fidelity. Our results
will have a significant impact on the ways physical resources are
distributed across a quantum network.
\end{abstract}
\maketitle

\section{Introduction}

Future quantum networks will allow one to exchange information over
large distances connecting multiple remote users \cite{NVnetworks1,NVnetworks2,Q_internet}.
This can be accomplished by sending high-quality quantum states, which
can then be used for a variety of tasks, for instance, improving the
security of the communication channels using quantum cryptographic
protocols \cite{QComm,Quantumcomm,QKD1,QKD2000,QKD02,QKD03,QC_Bill},
accelerating the computational time with quantum computers \cite{Qcomp1,Qcomp2,QComp_Bill,Quantum_comp2,Quantum_comp3,Quantum_comp4},
and improving the precision of measurements with quantum sensing and
imaging methods \cite{QSensing,QImaging3,QImaging2}. However, due
to the fragile nature of these quantum states errors and device imperfections
will affect the performance of those approaches cancelling the advantages
that these technologies have on their classical counterparts. 

One method that allows the transmission of high fidelity states involves
the use of quantum error correction (QEC) codes \cite{redundancy_code,surfacecode,GKP1,QEC2,QEC3,QEC4,QEC5,QEC1}.
Information is now encoded in a more complex quantum system, which
protect it from the errors occurring during transmission and recovered
when needed. The complexity of such code requires a large number of
physical resources for the encoding. Communication channels with low
capacities \cite{q_capacity1,q_capacity2,q_capacity3} and insufficient
resources within a node will reduce the number of resources that can
be transmitted over a single path, greatly affecting the communication
rate. For instance, when several users are connected by the same path
(or part of it), the number of channels of the path can be insufficient
for an efficient communication between two users, decreasing thus
their communication rate. Alternatively one can think of a single
channel connecting two users. The communication rate, in this case,
will be strictly limited by the repetition rate at which the photons
are sent.

One way to alleviate these issues is to connect the users with more
paths using quantum aggregation, in which the encoded states are distributed
over the channels of those distinct paths \cite{quantum_aggretation1}.
In \cite{quantum_aggretation1} it was shown that using two paths
for exchanging information using the quantum Reed-Solomon \cite{Reed_Sal}
(QRS) code leads to a drastic reduction of the transmittivity of the
channels of that path while increasing only slightly the transmittivity
of the other channels of the second path. Moreover, when higher-dimensional
photonic encodings are used \cite{Qmultiplexing} quantum aggregation
shows a drastic reduction of the physical resources required to reach
a threshold fidelity \cite{quantum_aggretation1}. However, in the
aggregation scenario a fundamental issue arises due to the different
length of the two paths. In fact, part of the encoded information
that arrives early at the remote site must be stored in a quantum
memory, which will undergo a dephasing process affecting the final
fidelity of the state \cite{reference_note2}. Once the delayed piece
of encoded state reaches the far end, it can be used with the one
retrieved from the quantum memory to correct the errors. The coherence
time of the quantum memories used can play therefore a fundamental
role in determining the performance of the QEC code when quantum aggregation
is in use. A too large difference in the length of the two paths or
a too short coherence time can be detrimental in recovering the information
sent making the communication among users impossible. In this work
we analyze the impact of temporal delays caused by the path length
differences (i.e., the time interval in which a piece of an encoded
quantum state is stored and that one in which it is retrieved) on
the fidelity of the final decoded state in a quantum aggregation scenario.
To this end, we consider two users that exchange information using
the QRS code connected by two and three communication paths of different
lengths. We determine the performance of such a system with delay
for several configurations in which the information can be distributed
and we determine the coherence times the quantum memories must have
for an optimal performance. 

The paper is divided as following: in Section II we analyze a quantum
aggregation system applied to the smallest QRS code with a temporal
delay in one path. Then in Section III we extend our analysis to higher
dimensional QRS codes and show several different and interesting configuration
arise. We conclude in Section IV. 

\section{quantum aggregation with delay}

\textcolor{black}{}
\begin{figure}
\begin{centering}
\includegraphics[scale=0.3]{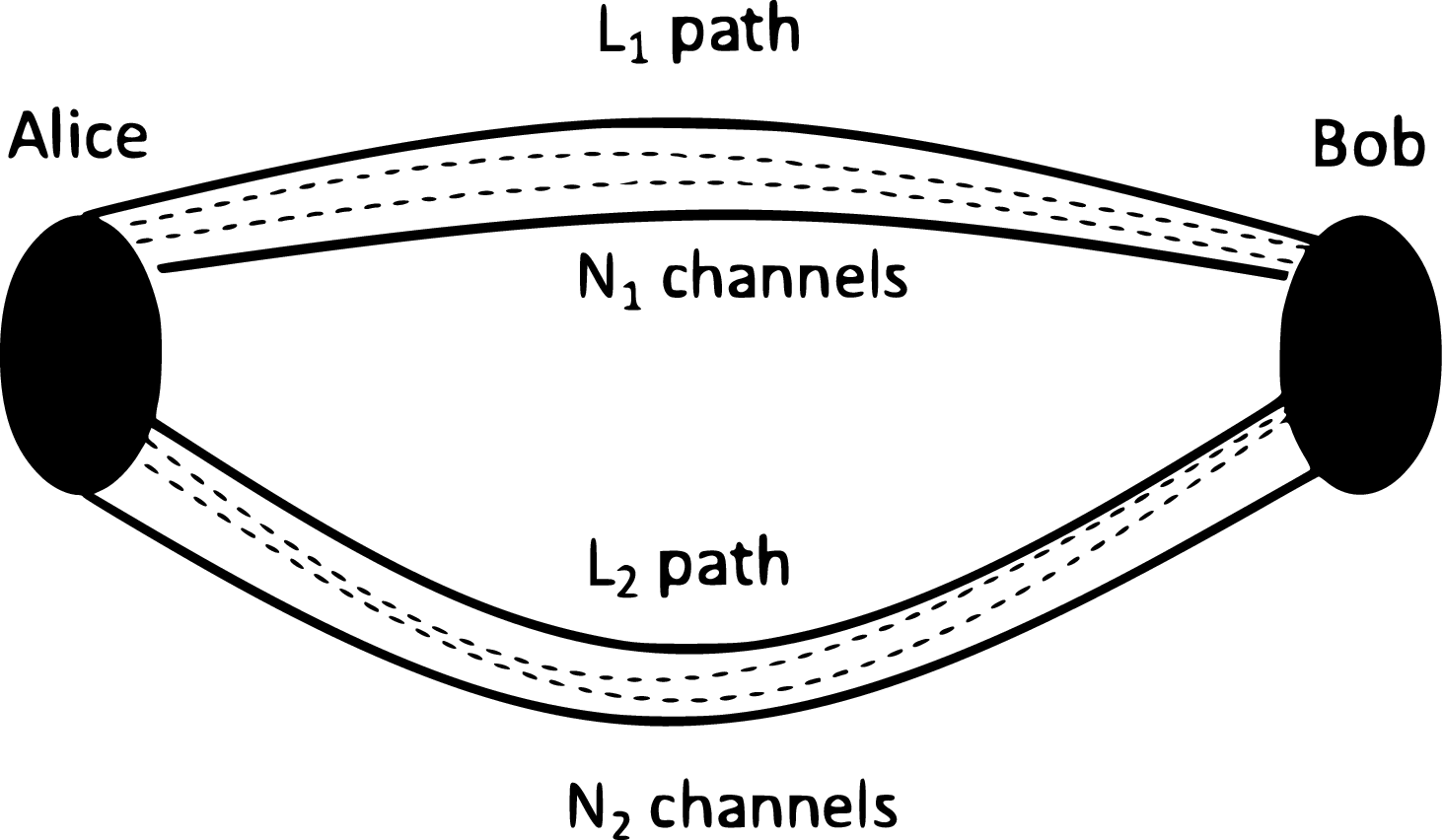}
\par\end{centering}
\textcolor{black}{\caption{Quantum aggregation over two paths of length $L_{2}>L_{1}$ containing
each $N_{1},$ $N_{2}$ channels, respectively. Alice distributes
her encoded state over the channels and send it through the two paths.
Bob stores the received early qudits into quantum memories and decodes
the state once the delayed qudits have arrived.}
}

\end{figure}
\textcolor{black}{Let us begin by exploring the effect of temporal
delay in quantum aggregation using the $[[n,1,d]]_{D}$ QRS code,
where $n$ is the number of physical qudits of dimension $D$ used
to encode one logical qudit and capable of correcting the loss of
$d-1$ qudits, with $d$ being the code distance. In the general quantum
aggregation scenario two users, Alice and Bob, are connected by two
lossy paths having different length, $L_{1}$ and $L_{2},$ as shown
in Fig. 1. In the following we assume that $D=n$ and $L_{2}>L_{1}$
with $N_{1}$ and $N_{2}$ being the number of channels inside path
1 and path 2, respectively \cite{reference_note2}. Alice encodes
her state using a $[[n,1,d]]_{D}$ QRS code and distributes $N_{1}$
qudits in the channels of path 1 and $N_{2}$ qudits in the channels
of path 2, respectively. We denoted such a configuration as $N_{1}+N_{2}.$
Then Bob decodes the received states if the number of the transmitted
qudits arriving earlier ($N_{1}^{'})$ is enough for retrieving the
information sent by Alice ($N_{1}^{'}\geq d)$, otherwise, when $N_{1}^{'}<d$,
he stores those qudits in quantum memories (QMs). We assume that the
density matrix, $\rho,$ of these stored qudits undergo a depolarizing
channel given by $\rho\rightarrow\rho'=(1-p_{d})\rho+p_{d}I/D',$
where $I$ is the $D'$ dimensional identity operator and $p_{d}$
is the depolarization error probability given by $p_{d}=1-e^{-t/T_{2}},$
with $t=\left(L_{2}-L_{1}\right)/c$ and $T_{2}$ being the coherence
time of the QMs. Next when $N_{1}^{'}<d$ while the number of qudits
transmitted over path 2 ($N_{2}^{'})$ satisfies $N_{2}^{'}\geq d,$
Bob uses the $N_{2}^{'}$ qudits to recover the initial information
discarding the stored qudits associated with the transmission through
path $L_{1}.$ Now when both $N_{1}^{'}<d$ and $N_{2}^{'}<d$ but
with $N_{1}^{'}+N_{2}^{'}\geq d$ Bob, retrieves the qudits stored
into the QMs and applies a decoding procedure on all transmitted qudits.
This latter case will affect the fidelity of the decoded state due
to the $temporal$ $delay$ of the qudits traveling in path 2. Finally,
when $N_{1}^{'}+N_{2}^{'}<d$ we assume for simplicity the state shared
by Alice and Bob is a completely mixed state (the worst case) and
all the information has been lost. We assume that the local gates
errors are negligible compared to the memory depolarization errors.}

Now let us explore the impact of the temporal delay in a quantum aggregation
scenario using the smallest QRS code the $[[3,1,2]]_{3}$ code capable
of correcting one error in which one logic qutrit is created using
three physical qutrits. In the $[[3,1,2]]_{3}$ QRS code protocol
Alice encodes her initial qutrit $\left.|\psi\right\rangle _{A}=\left.\alpha_{0}|0\right\rangle +\left.\alpha_{1}|1\right\rangle +\left.\alpha_{2}|2\right\rangle $
into the logic state $\left.|\psi\right\rangle _{L}=\left.\alpha_{0}|0\right\rangle _{L}+\left.\alpha_{1}|1\right\rangle _{L}+\left.\alpha_{2}|2\right\rangle _{L},$
where $\left.|0\right\rangle _{L}=\left(\left.|000\right\rangle +\left.|111\right\rangle +\left.|222\right\rangle \right)/\sqrt{3};$
$\left.|1\right\rangle _{L}=\left(\left.|012\right\rangle +\left.|120\right\rangle +\left.|201\right\rangle \right)/\sqrt{3}$
and $\left.|2\right\rangle _{L}=\left(\left.|021\right\rangle +\left.|102\right\rangle +\left.|210\right\rangle \right)/\sqrt{3}.$
She sends this encoded state over a lossy path to Bob. Upon a successful
transmission of the state sent by Alice, Bob applies a decoding procedure
described in \cite{decoding_LJ} to retrieve the initial state. In
the quantum aggregation scenario, we have two configurations; the
$2+1$ configuration and the $1+2$ configuration, in which $2(1)$
qudits are traveling in the channels of path 1 having a transmissivity
$p_{1}=e^{-L_{1}/L_{\mathrm{att}}}$ while $1(2)$ qutrits are sent
via path 2 with transmission probability $p_{2}=e^{-L_{2}/L_{\mathrm{att}}},$
respectively. Here $L_{\mathrm{att}}=22$ km is the attenuation length
of the optical fiber channels.

\begin{figure}
\begin{centering}
\includegraphics[scale=0.17]{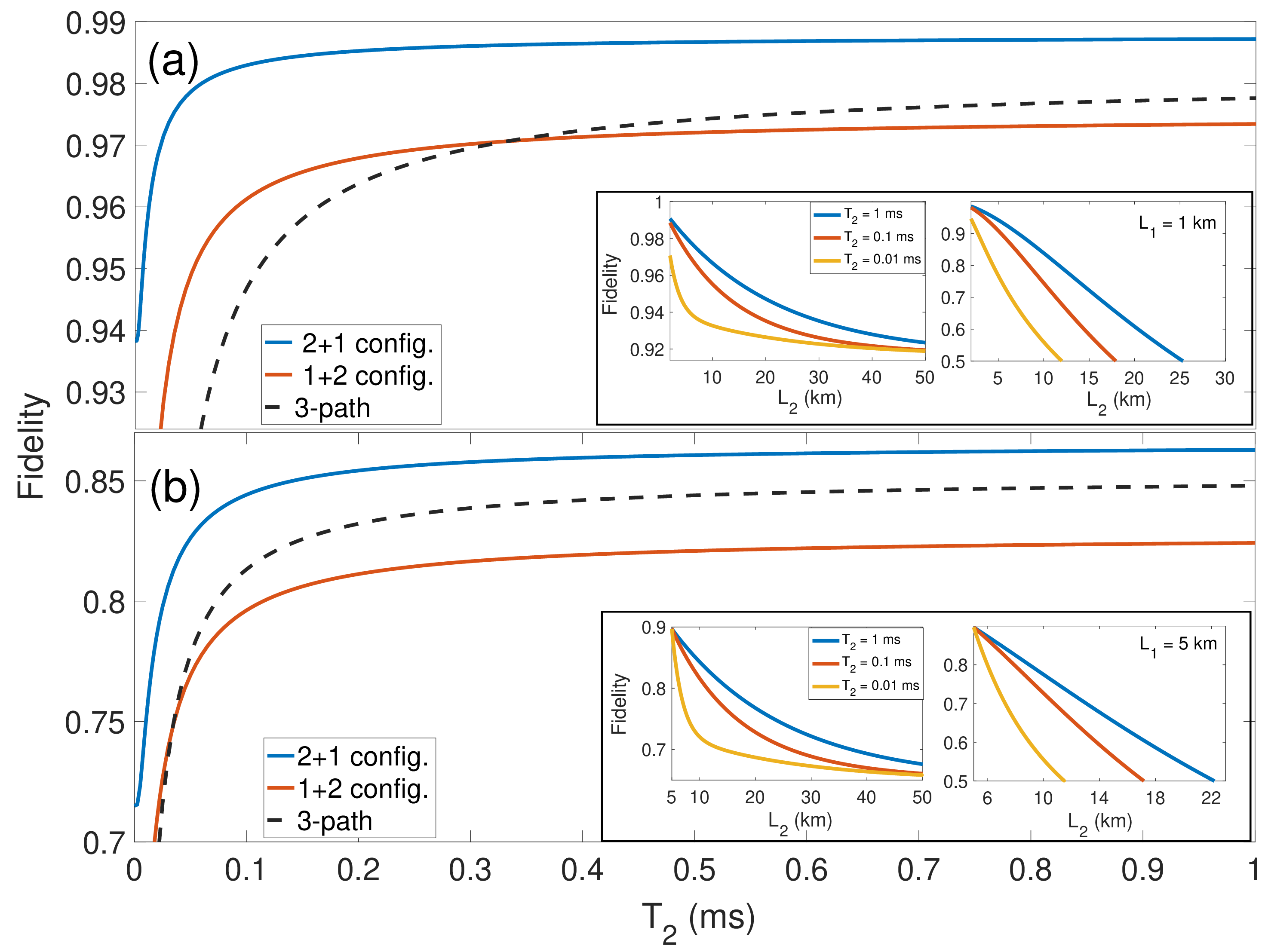}
\par\end{centering}
\caption{\label{fig:Fidelity-of-state}Fidelity of the decoded state received
by Bob versus the coherence time, $T_{2},$ of the QM used to store
the qutrits transmitted through path 1, in the 2+1 configuration (solid
blue lines) and 1+2 configuration (solid red lines) for (a) $L_{1}=1$
km, $L_{2}=3$ km and (b) $L_{1}=5$ km, $L_{2}=8$ km. In the insets
we depict the fidelity versus $L_{2}$ at (a) $L_{1}=1$ km and (b)
$L_{1}=5$ km for both configurations. Also shown for the 3-path configuration
as the black dashed curves (a) with $L_{1}=1$ km, $L_{2}=2$ km,
$L_{3}=3$ km and (b) with $L_{1}=5$ km, $L_{2}=6$ km, $L_{3}=8$
km.}
\end{figure}

In the 2+1 configuration the fidelity, $F_{2+1},$ of the state received
and decoded by Bob is

\begin{equation}
\begin{aligned}F_{2+1} & =p_{1}^{2}p_{2}+2p_{1}p_{2}(1-p_{1})\left(1-\frac{2}{3}p_{d}\right)\\
 & +p_{1}^{2}(1-p_{2})+(1-P_{s_{1}})/27
\end{aligned}
\label{eq:3qd_2chs}
\end{equation}
where $P_{s_{1}}=p_{1}^{2}p_{2}+2p_{1}p_{2}(1-p_{1})+p_{1}^{2}(1-p_{2})$
is the probability of the successful transmission of information from
Alice to Bob. Let us give an intuitive derivation of Eq. \eqref{eq:3qd_2chs}
that can be easily extended to derive the fidelity of higher dimensional
QRS codes. The no-loss term, which is the first and dominant term
in Eq. \eqref{eq:3qd_2chs}, and the loss of the qudit traveling in
path 2 (third term in Eq. \eqref{eq:3qd_2chs}) do not depend on the
depolarization error $p_{d}$ because Bob applies immediately the
decoding procedure on the two qutrits traveling in the channels of
path 1. Then, in the case in which one of the two qudits traveling
in path 1 is lost, the temporal delay due to the storage of the transmitted
qudit contributes to the fidelity with a term proportional to $\left(1-\frac{2}{3}p_{d}\right),$
which has been derived in Appendix A. To gain an understanding of
the behavior of the fidelity we plot in Fig. \ref{fig:Fidelity-of-state}(a)
the fidelity $F_{2+1}$ versus the memories coherence time for $L_{1}=1$
km and $L_{2}=3$ km (solid blue curve). We see that the coherence
time of the QM only affects the fidelity when $T_{2}<0.1$ ms and
noting that for no memory $(T_{2}=0)$ $F_{2+1}\sim0.94.$ This is
due to the fact that the the no-loss term does not depend on $p_{d}.$
Therefore, even when there are no QMs in the system the transmitted
state can be used to extract some information. A similar explanation
can be given to the case in which $L_{2}\rightarrow\infty,$ (see
the left graph of the inset of Fig. \ref{fig:Fidelity-of-state}(a)).
Here we plot the fidelity versus $L_{2}$ with $L_{1}=1$ km for $T_{2}=$
1 ms (blue curve), $T_{2}=$ 0.1 ms (red curve) and $T_{2}=0.01$
ms, respectively. We observe that the fidelity decreases at lower
coherence times while reaching an asymptotic value of $\sim0.92$
at large values of $L_{2},$ from which information can partially
be extracted. This can be explained considering that, for this configuration,
the no-loss term corresponds to the case in which two qudits are successfully
transmitted and one is lost with very high probability. Therefore,
since this code can correct the loss of one qudit, the transmitted
state containing two qudits with probability ($\sim p_{1}^{2})$,
still has sufficient information sallowing the fidelity to exceed
$50\%.$

The alternate 1+2 configuration changes quite drastically. It is straight
forward to show that $F_{1+2}$ is 

\begin{equation}
\begin{aligned}F_{1+2} & =p_{2}^{2}p_{1}+2p_{1}p_{2}(1-p_{2})\left(1-\frac{2}{3}p_{d}\right)\\
 & +p_{2}^{2}(1-p_{1})+(1-P_{s_{2}})/27
\end{aligned}
\label{eq:1+2 fid}
\end{equation}
where $P_{s_{2}}=p_{2}^{2}p_{1}+2p_{1}p_{2}(1-p_{2})+p_{2}^{2}(1-p_{1}).$
Even in this configuration the dominant term is the no-loss term,
which does not depend on $p_{d}$ because, although the two qudits
arrive later, they can be immediately be used to decode the state
while discarding the early qudit transmitted over the path 1. The
second term in Eq. \eqref{eq:1+2 fid} refers to the lost of a qudit
traveling in path 2. Therefore, Bob needs to retrieve the stored qudit
from the QM to decode the state together with the single qudit transmitted
over path 2. The contribution to the fidelity from the depolarization
channel applied to the stored qudit is equal to the previous configuration.
Finally, the third term in Eq. \eqref{eq:1+2 fid} does not depend
on $p_{d}$ because Bob can use the two qudits transmitted over path
2 to decode the state. To visualize this we plot $F_{2+1}$ (solid
red line) versus $T_{2}$ at the same numerical values of the previous
configuration, as shown in Fig. \ref{fig:Fidelity-of-state}(a). As
expected, the fidelity in this case has much lower values because
the no-loss term suffers the loss of two qudits with higher probability,
hence the second term in Eq. \eqref{eq:1+2 fid} is more relevant
in this case. In other words the coherence time in this case affect
the fidelity more than the other case. This can also be seen from
the graph in the right side of the inset of Fig. \ref{fig:Fidelity-of-state}(a).
Here, we plot the fidelity versus $L_{2}$ for different coherence
times. In this case the probability of losing two qudits traveling
in path 2 increases with $L_{2},$ (mathematically, the first term
in Eq. \eqref{eq:1+2 fid} decreases), hence, the contribution to
the fidelity from the second term is more significant. In this case
one can see that $T_{2}$ determines a threshold value for $L_{2}$
after which the fidelity is below $50\%$ (see the crossing points
of the curves with the $x$ axis in the graph in the right side of
the inset of Fig. \ref{fig:Fidelity-of-state}(a)). From this considerations
we conclude (as expected) that distributing more qudits in the shorter
channel gives a significant advantage in terms of having higher fidelities
and being slightly affected by the coherence time. Please see Appendix
B for the full derivation of Eq. \eqref{eq:3qd_2chs} and \eqref{eq:1+2 fid}.

We expect that for higher values of $L_{1}$ all the results described
above are worse for both configurations. This scenario is shown in
Fig. \ref{fig:Fidelity-of-state}(b), where we plot the fidelities
of the configurations versus $T_{2}$ at $L_{1}=5$ km and $L_{2}=$
8 km and in the inset where we plot the fidelities of both configurations
versus $L_{2}$ at $L_{1}=5$ km. Even in this case the fidelity in
the $2+1$ configuration reaches an asymptotic value as $L_{2}$ increases,
which is much lower than the previous case. 

It is interesting now to add one more path (path 3), with transmission
probability $p_{3}=e^{-L_{3}/L_{\mathrm{att}}},$ to the previous
scheme, such that $L_{3}>L_{2}>L_{1}$ while maintaining the same
highest distance separating Alice and Bob. In this case, each qutrit
travels across a channel in the corresponding path. This can be referred
to as the 1+1+1 configuration. The fidelity of Bob's decoded state
is

\begin{equation}
\begin{array}{c}
\begin{aligned}F_{1+1+1} & =p_{1}p_{2}p_{3}\left(1-\frac{2}{3}p_{d_{12}}\right)\\
 & +p_{2}p_{3}(1-p_{1})\left(1-\frac{2}{3}p_{d_{23}}\right)\\
 & +p_{1}p_{3}(1-p_{2})\left(1-\frac{2}{3}p_{d_{13}}\right)\\
 & +p_{1}p_{2}(1-p_{3})\left(1-\frac{2}{3}p_{d_{12}}\right)\\
 & +(1-P_{s})1/27,
\end{aligned}
\end{array}\label{eq:3qds_3chs}
\end{equation}
where $p_{d_{ij}}=1-e^{-T_{ij}/T_{2}},$ with $T_{ij}=|L_{i}-L_{j}|/c$
and $P_{s}=p_{1}p_{2}p_{3}+p_{2}p_{3}(1-p_{1})+p_{1}p_{3}(1-p_{2})+p_{1}p_{2}(1-p_{3}).$
In this case, the no-loss term of Eq. \eqref{eq:3qds_3chs} depends
on $p_{d}$ because the qutrit traveling in the channel of path 1
arrives first and needs to be stored in a QM before Bob can apply
a decoding process with a second qudit. One can also see that all
the other terms in Eq. \eqref{eq:3qds_3chs} depends on $p_{d}$ because
in the loss event of any qudit, Bob needs to wait for another one
to start the decoding process. Figure \ref{fig:Fidelity-of-state}(a)
shows the fidelity of Eq. \eqref{eq:3qds_3chs} (black dashed line)
at $L_{1}=1$ km, $L_{2}=2$ km and $L_{3}=3$ km. One can see that
for small values of $T_{2}$ ($T_{2}<0.1$ ms) the no loss term greatly
affects the fidelity whereas for higher values of $T_{2}$ the fidelity
of the 3-path case increases until it crosses the 1+2 configuration
of the 2-path case at a crossing point $T_{2}^{c}\simeq0.3$ ms. This
is due to the fact that the expression of the fidelity of the state
received by Bob in the 3-path case is more affected by the coherence
time than $1+2$ case as one can see comparing Eq. \eqref{eq:3qd_2chs}
with Eq. \eqref{eq:3qds_3chs}. Therefore we expect that for large
values of $T_{2}$ the no-loss term of Eq. \eqref{eq:3qds_3chs} becomes
higher than the no-loss term of Eq. \eqref{eq:3qd_2chs} because when
$T_{2}\rightarrow\infty$ the qudit in path 2 travels over a smaller
distance than the qudits of the 1+2 configuration. At $L_{1}=5$ km
and $L_{2}=8$ km the 3-path case has also a lower fidelity as shown
from the dashed curve in Fig. \ref{fig:Fidelity-of-state}(b). However,
in this case the crossing point of this curve with the one corresponding
to the $1+2$ configuration is slightly lower than the crossing point
shows in Fig. \ref{fig:Fidelity-of-state}(a). This can be explained
by considering that for larger distances the main source of error
is the channel loss, hence, the coherence time affects less the fidelity
of the decoded state. In fact we can see that at very low values of
$T_{2}$ the fidelity of the 3-path case is very similar to the $1+2$
case. 

\section{temporal delay for higher dimensional codes}

So far we have considered the smallest QRS code that can correct 1
loss errors. What happens as we increase the code size? We analyze
the effects of the temporal delay in a quantum aggregation scenario
for the $[[5,1,3]]_{5}$ and the $[[7,1,4]]_{7},$ QRS codes, which
can fix the loss of 2 and 3 qudits, respectively. Let us begin with
the $[[5,1,3]]_{5}$ code. Here there are 4 possible configurations:
4+1, 3+2, 2+3 and 1+4.

\subsection{The 4+1 and 1+4 configurations}

In these configurations for the $[[5,1,3]]_{5}$ code, the fidelity
of the state decoded by Bob is:

\begin{equation}
\begin{array}{c}
\begin{aligned}F_{4+1} & =p_{1}^{4}p_{2}+4p_{1}^{3}p_{2}(1-p_{1})+p_{1}^{4}(1-p_{2})+\\
 & +6p_{1}^{2}p_{2}(1-p_{1})^{2}f_{1}(p_{d})\\
 & +4p_{1}^{3}(1-p_{1})(1-p_{2})+(1-P_{s_{1}})/5^{5}
\end{aligned}
\end{array}\label{eq:Fid5_L1mL2}
\end{equation}
and 

\begin{equation}
\begin{array}{c}
\begin{aligned}F_{1+4} & =p_{2}^{4}p_{1}+4p_{2}^{3}p_{1}(1-p_{2})+p_{2}^{4}(1-p_{1})\\
 & +6p_{2}^{2}p_{1}(1-p_{2})^{2}f_{2}(p_{d})\\
 & +4p_{2}^{3}(1-p_{2})(1-p_{1})+(1-P_{s_{2}})/5^{5}.
\end{aligned}
\end{array}\label{eq:Fid5L2mL1}
\end{equation}
where $P_{s_{1(2)}}=p_{1(2)}^{4}p_{2(1)}+4p_{1(2)}^{3}p_{2(1)}(1-p_{1(2)})+p_{1(2)}^{4}(1-p_{2(1)})+6p_{1(2)}^{2}p_{2(1)}(1-p_{1(2)})^{2}+4p_{1(2)}^{3}(1-p_{1(2)})(1-p_{2(1)})$
and $f_{1(2)}(p_{d})$ being a contribution to the fidelity when 2
or 1 qudits are dephasing, respectively.

Comparing Eq. \eqref{eq:Fid5_L1mL2} with Eq. \eqref{eq:Fid5L2mL1}
one can see that the expressions of the two fidelities are almost
identical except for the 4th term, which is multiplied by $f_{1,2}(p_{d}),$
respectively, whose analytical expression is given in Appendix B.
What however is important is that $f_{1}(p_{d})\leq f_{2}(p_{d})$
and only equal at $p_{d}=0,$ $1.$ This is due to the fact that the
term $f_{1}(p_{d})$ can be considered as the fidelity of a density
matrix in which 2 qudits are dephasing whereas $f_{1}(p_{d})$ takes
into account the dephasing of a single qudit. Hence in this latter
case, less information has been lost. However, the dominant term in
both Eq. \eqref{eq:Fid5_L1mL2} and Eq. \eqref{eq:Fid5L2mL1} is the
no-loss term, hence, 
\begin{figure}
\begin{centering}
\includegraphics[scale=0.14]{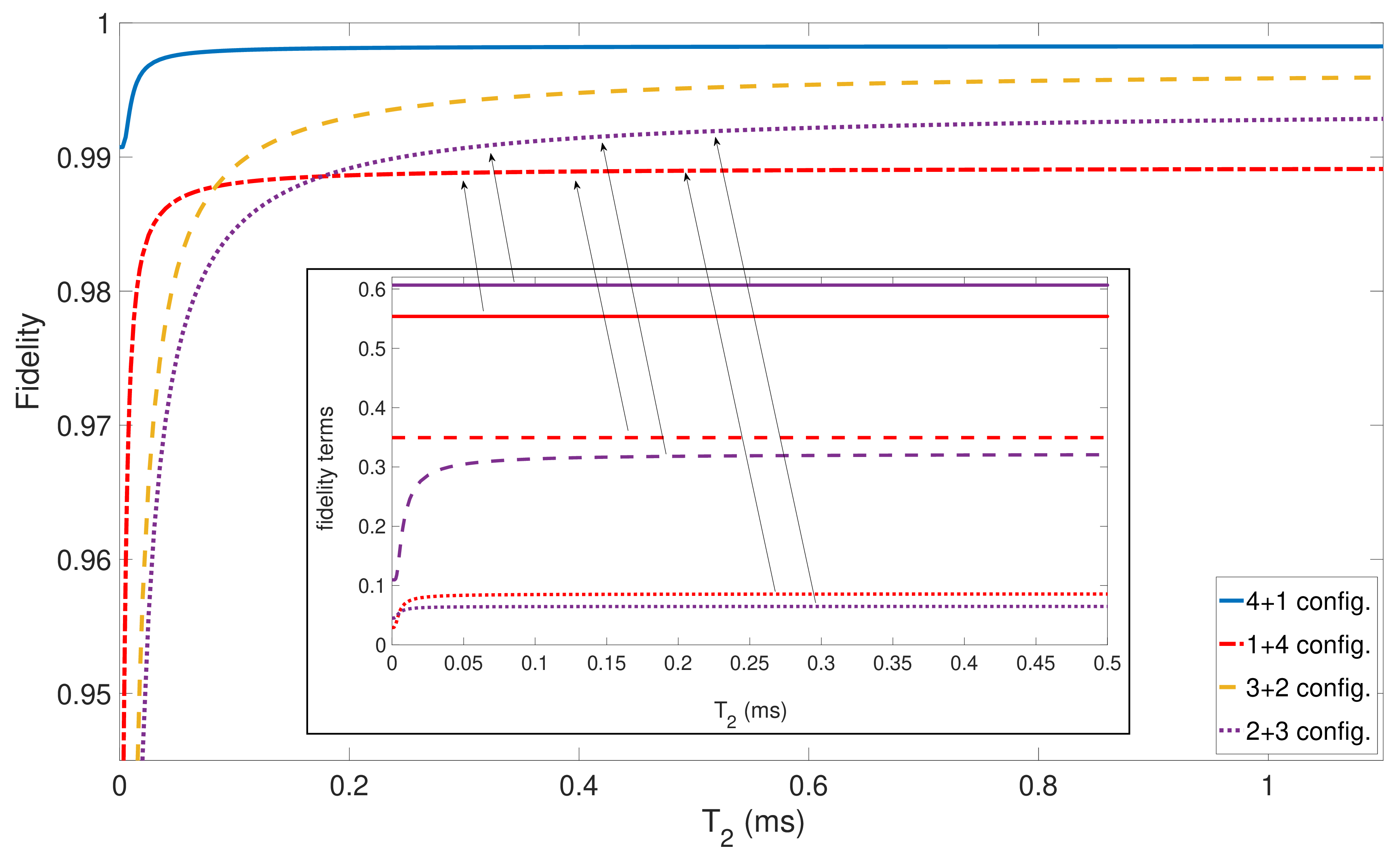}
\par\end{centering}
\caption{\label{fig:fidelity_5d}Fidelity of the state decoded by Bob when
Alice encodes her state using the $[[5,1,3]]_{5}$ QRS code in a quantum
aggregation scenario for $L_{1}=$ 1 km and $L_{2}=3$ km. The blue
solid curve (red dash-dotted curve) refers to the 4+1 (1+4) configuration
in which Alice distribute four (one) qudits into the shorter channels
and one (four) qudits in the longer channels, respectively, whereas
the dashed yellow curve (dotted purple curve) refers to the 3+2 configuration
in which Alice distributes three (two) qudits into the shorter channels
and three(two) qudits in the longer ones respectively. In the inset,
we plot the contributions to the fidelity of the probability of losing
zero (solid curves), one (dashed curves) and two (dotted curves) qudits
for the 1+4 and 2+3 configurations.}
\end{figure}
 $F_{1}$ is higher than $F_{2}$ for any value of the dephasing time
$T_{2},$ as shown in Fig. Fig. \ref{fig:fidelity_5d}. Therefore
it is more advantageous to distribute more qudits into the shorter
path as well as we obtained for the three dimensional code.

\subsection{The 3+2 and 2+3 configurations}

It is now interesting to analyze the 3+2 and 2+3 configurations of
the $[[5,1,3]]_{5}$ QRS code. The respective fidelities are

\begin{equation}
\begin{array}{c}
\begin{aligned}F_{3+2} & =p_{1}^{3}p_{2}^{2}+3p_{1}^{2}p_{2}^{2}(1-p_{1})f_{3}(p_{d})+2p_{1}^{3}p_{2}(1-p_{2})\\
 & +3p_{1}p_{2}^{2}(1-p_{1})^{2}f_{2}(p_{d})+p_{1}^{3}(1-p_{2})^{2}\\
 & +6p_{1}^{2}p_{2}(1-p_{1})(1-p_{2})f_{1}(p_{d})+(1-P_{s_{1}})/5^{5}
\end{aligned}
\end{array}\label{eq:fid3p2_L1mL2}
\end{equation}
and

\begin{equation}
\begin{array}{c}
\begin{aligned}F_{2+3} & =p_{2}^{3}p_{1}^{2}+3p_{2}^{2}p_{1}^{2}(1-p_{2})f_{3}(p_{d})+2p_{2}^{3}p_{1}(1-p_{1})\\
 & +3p_{2}p_{1}^{2}(1-p_{2})^{2}f_{1}(p_{d})+p_{2}^{3}(1-p_{1})^{2}\\
 & +6p_{2}^{2}p_{1}(1-p_{2})(1-p_{1})f_{2}(p_{d})+(1-P_{s_{2}})/5^{5},
\end{aligned}
\end{array}\label{eq:Fid3p2_L2mL1}
\end{equation}
where $P_{s}=p_{1(2)}^{3}p_{2(1)}^{2}+3p_{1(2)}^{2}p_{2(1)}^{2}(1-p_{1(2)})+2p_{1(2)}^{3}p_{2(1)}(1-p_{2(1)})+3p_{1(2)}p_{2(1)}^{2}(1-p_{1(2)})^{2}+p_{1(2)}^{3}(1-p_{2(1)})^{2}+6p_{1(2)}^{2}p_{2(1)}(1-p_{1(2)})(1-p_{2(1)}).$ 

Figure \ref{fig:fidelity_5d} shows that, even in this configuration,
it is more convenient to use more qudits in the shorter path. In fact,
$F_{1}$ (dashed yellow curve) is higher than $F_{2}$ (dotted purple
curve) for any value of $T_{2}.$ Further, Fig. \ref{fig:fidelity_5d}
shows that at $T_{2}\geq0.16$ ms the fidelity of the $2+3$ configuration
outperforms the fidelity of the $1+4$ configuration. This can be
explained with the fact that some terms of Eq. \eqref{eq:Fid3p2_L2mL1}
are much more affected by the dephasing channel than Eq. \eqref{eq:Fid5L2mL1}.
In fact, in the inset of Fig. \ref{fig:fidelity_5d} we plot for those
two configurations, the contributions to the fidelity coming from
the probability of losing zero (solid curves), one (dashed curves)
and two (dotted curves) qudits. We expect that that for large values
of $T_{2}$ the probability of not losing any qudit is higher in the
2+3 configuration (purple solid curve of the inset) because less qudits
are traveling in the longer channel compared to the 1+4 configuration
(red solid curve of the inset) whereas the loss terms must be smaller.
On the other hand, at lower values of $T_{2}$, the contribution of
losing one qudit for the 2+3 configuration is strongly affected by
dephasing whereas the one of the 1+4 configuration does not depend
on it. As regards the probability of losing two qudits, the dephasing
channel affects both configurations with the 2+3 being slightly lower. 

\subsection{The $[[7,1,4]]_{7}$ QRS code}

\begin{figure}
\begin{centering}
\includegraphics[scale=0.16]{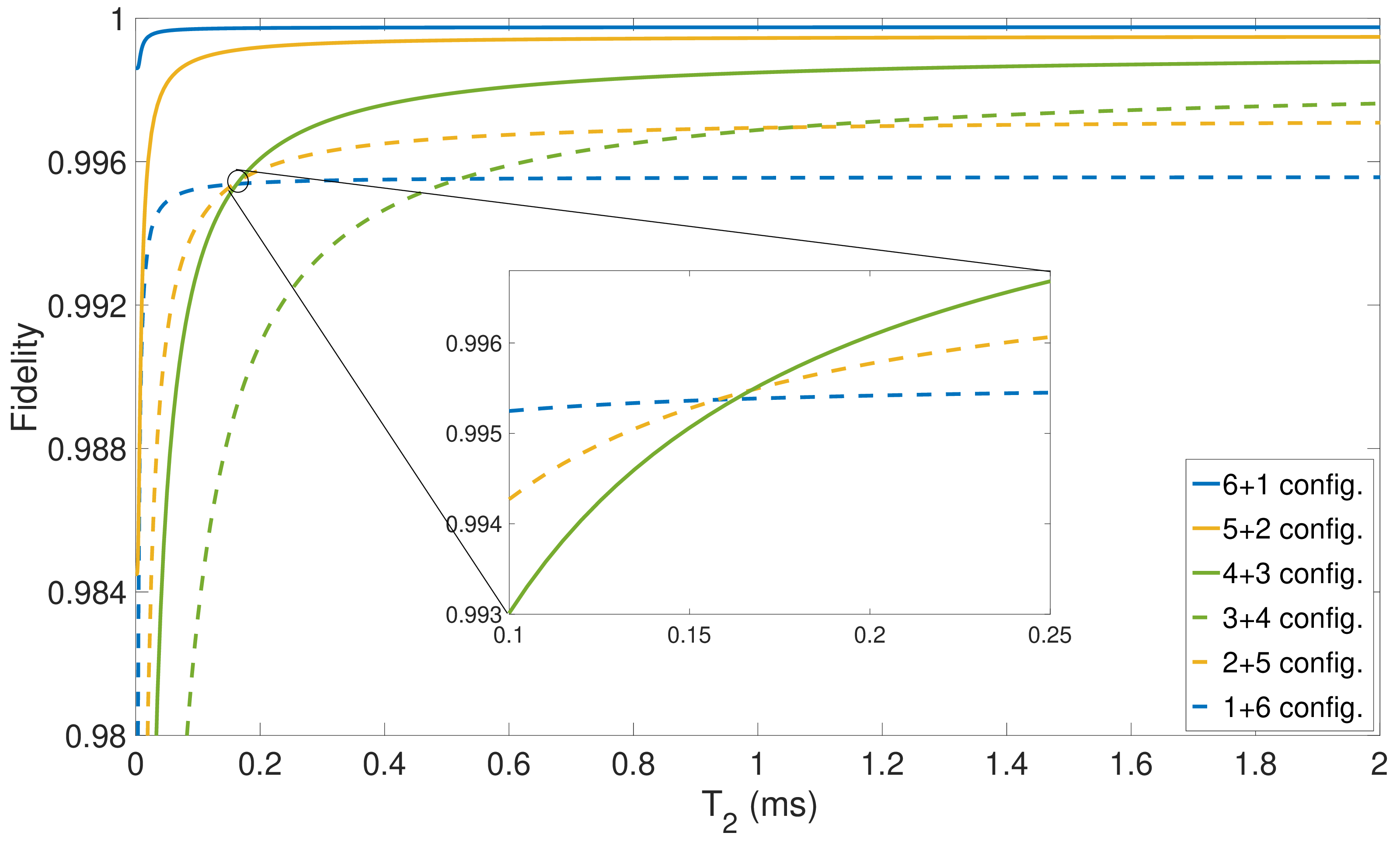}
\par\end{centering}
\caption{\label{fig:7_dim_graph}Fidelity of the state received by Bob when
Alice encodes her state using the $[[7,1,4]]_{7}$ QRS code in a quantum
aggregation scenario for $L_{1}=1$ km and $L_{2}=3$ km. The blue
solid curve (blue dashed curve) refers to the 6+1 (1+6) configuration
in which Alice distribute six (one) qudits into the shorter channels
and one (six) qudits in the longer channels, respectively; the solid
yellow curve (dashed yellow curve) refers to the 5+2 configuration
in which Alice distribute five (two) qudits into the shorter channels
and two (five) qudits in the longer ones, respectively; the solid
green curve (dashed green curve) refers to the 4+3 configuration in
which Alice distribute four (three) qudits into the shorter channels
and three (four) qudits in the longer ones, respectively. In the inset
we plot the portion of the graph in which the 1+6, 2+5 and 4+3 configurations
cross.}
\end{figure}
In this subsection we show the results of the $[[7,1,4]]_{7}$ QRS
code. The analytical expression of the fidelities for all possible
configurations are listed in Appendix C. Figure \ref{fig:7_dim_graph}
shows the fidelities of all configurations of the $[[7,1,4]]_{7}$
QRS code, where the solid curves refer to the case in which a higher
number of qudits travel in the shorter channels whereas the dashed
curves refer to the case in which the smaller number of qudits travel
in longer channels. One can notice the following common features shared
with the $[[5,1,3]]_{5}$ QRS case illustrated above. Firstly the
case in which more qudits travel in the shorter path has higher fidelity
than the other case for any value of $T_{2}.$ This is mainly due
to the contribution of the no-loss term, which is the dominant term
in the expressions of the fidelities. Then, when the qudits are almost
equally distributed between the two channels (for instance the 3+2
configuration of the $[[5,1,3]]_{5}$ QRS code, or the 4+3 configuration
of the $[[7,1,4]]_{7}$ QRS code) the fidelities are much more affected
by the dephasing. As a consequence, these fidelities will reach the
asymptotic limit of $T_{2}\rightarrow\infty$ at higher values of
$T_{2}$ as shown in Fig. \ref{fig:7_dim_graph} (for instance, the
purple curve in Fig. \ref{fig:fidelity_5d} and the green curves in
Fig. \ref{fig:7_dim_graph}). Then, as well as the this case of the
$[[5,1,3]]_{5}$ QRS code, there is a crossing point in which the
fidelities of different configurations intersect. The inset of Fig.
\ref{fig:7_dim_graph} shows that this dephasing crossing point, $T_{2}^{c},$
occurs at $T_{2}^{c}=0.16$ ms. This can be explained with a very
similar motivation given in the $[[5,1,3]]_{5}$ QRS code case. In
fact, the loss terms in the fidelity's expressions of the configurations
in which the qudits are more evenly distributed depend on the dephasing
channel much more than the loss terms of the uneven distributions,
as one can see in Eqs. \eqref{eq:2+5} - \eqref{eq:3+4} compared
to Eq. \eqref{eq:1+6}. Hence, the corresponding fidelities assume
high values at $T_{2}\rightarrow\infty$ and very low values at $T_{2}\rightarrow0$
for the even distribution cases leading to a crossing point with the
fidelity of the uneven case. \textcolor{black}{This is an interesting
feature of the aggregation network because one user can achieve a
faster communication rate, sending more qudits simultaneously the
more even is the distribution, while having better fidelities than
the more uneven distribution case for certain values of $T_{2}.$
On the other hand, when the highest value of fidelity is required,
then a more uneven distribution is preferred. This aspect can play
an important role for some quantum communication systems in which
a trade-off between fidelity of the transmitted states and transmission
rate is the key factor, such as in several quantum key distribution
schemes \cite{QKD1,QKD02,QKD03,QKD2000}.}\textcolor{red}{{} }

\section{Conclusion and discussion}

Distributing physical resources over multiple channels of different
length in a quantum aggregation scenario will require the use of quantum
memories to store the states arriving earlier. The decoherence process
occurring in the memories will partially destroy the stored information
before the delayed state arrives. Here we analyze the effect of such
a delay time in a QRS code having dimension three, five and seven,
respectively. For these codes, we analytically calculate the fidelity
of the final state as function of the channel loss and the dephasing
time for different configurations in which the resources are evenly
or unevenly distributed over two paths of different length. We obtain
that for a coherence times $T_{2}>1$ ms the fidelities of all the
configurations asymptotically reach their optimal value. This threshold
for the coherence time is vastly reachable with today's technology
using, for instance, ion qubits \cite{ion_qubit}, superconducting
cavities \cite{supercond_cavity}, nuclear qubits of NV centers \cite{NVKae}
and ensemble-based quantum memories \cite{ensemble_based}. We also
analyze the behavior of such fidelities at a fixed value of the coherence
time when the length difference between the two paths increases. In
this case we obtain an asymptotic value for the fidelity when the
majority of the qudits travels across the shorter path regardless
the value of the coherence time. 

On the other hand, when most of the qudits travel in the longer path
we determine the largest achievable distance of such a path, which
is strongly affected by the coherence time. We show that while quantum
aggregation allows users in a quantum network to exchange information
faster, the impact of a temporal delay in the received states can
have a detrimental effect on the quality of the transmitted information.
The secret key bit rate can be a good figure of merit to estimate
the performance of a quantum network since it takes into account both
the repetition rate at which bits are shared between two remote parties
as well as the quality of the density matrix shared by them. Optimizing
the secret key rate using quantum aggregation can therefore be a valid
route to follow for the evaluation of the performance of tomorrow's
quantum networks. Besides the configurations analyzed in this work
can potentially provide a guideline on the architecture of quantum
networks. Future works might consider using other error correction
codes with quantum aggregation due to its versatility as well as adding
more paths connecting users.
\begin{acknowledgments}
\textcolor{black}{This project was made possible through the support
of the Moonshot R\&D Program Grants JPMJMS2061 \& JPMJMS226C and JSPS
KAKENHI Grant No. 21H04880.}

\end{acknowledgments}

\bibliographystyle{apsrev4-1}
\bibliography{bib1}

\appendix

\section{Decoding procedures}

In this Appendix we illustrate the procedure to recover the state
sent by Alice for the $[[5,1,3]]_{5}$ and $[[7,1,4]]_{7}$ QRS codes,
respectively, in a lossy channel. Figure \ref{fig:dec_5d} shows the
circuit that Bob applies to the encoded state sent by Alice when (a)
illustrates the situation involving the loss of a single qudit and
(b) the loss of two qudits. The gates represent a sum modulo $5$
between two qudits of dimension $5.$ After these gates are applied
the remaining qudits except the first one are measured in the computational
basis. These measurements will ideally project the first qudit into
the state $\left.|\psi\right\rangle _{A}=\alpha_{0}\left.|0\right\rangle +\alpha_{1}\left.|1\right\rangle +\alpha_{2}\left.|2\right\rangle +\alpha_{3}\left.|3\right\rangle +\alpha_{4}\left.|4\right\rangle .$
Similarly, it is possible to retrieve the initial state of Alice for
the $[[7,1,4]]_{7}$ QRS code with the decoding circuits shown in
Fig. \ref{fig:dec_7d}(a, b, c), which correspond to the loss of one,
two and three qudits, respectively.

\section{Fidelities after dephasing}

\begin{figure}
\begin{centering}
\includegraphics[scale=0.3]{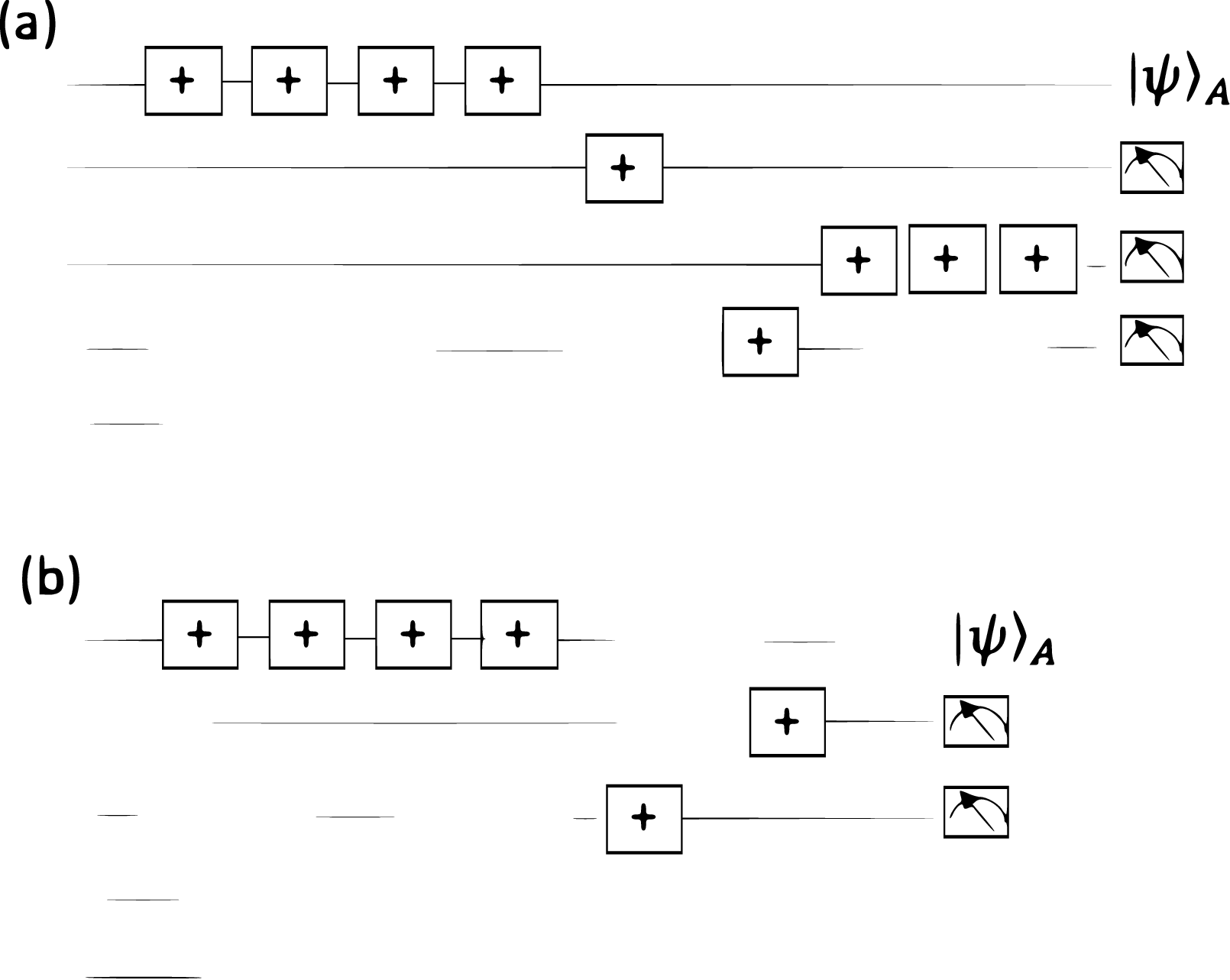}
\par\end{centering}
\caption{\label{fig:dec_5d}Decoding circuit of the $[[5,1,3]]_{5}$ QRS code
when (a) one qudit and (b) two qudits are lost, respectively. The
dashed lines refer to the loss of a qudit. The symbol ``+'' refers
to the sum mod 5 gate.}
\end{figure}
\begin{figure}
\begin{centering}
\includegraphics[scale=0.27]{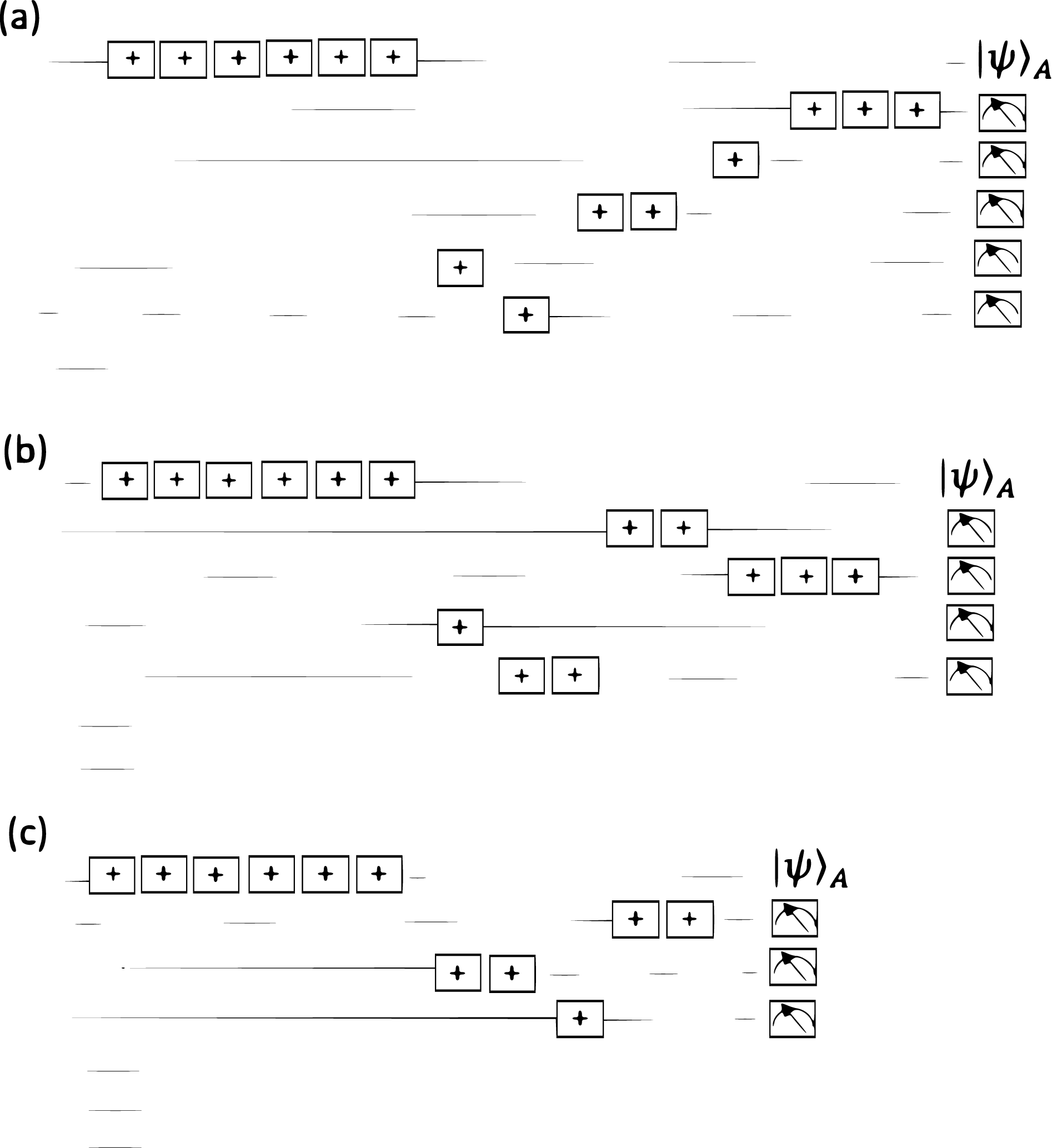}
\par\end{centering}
\caption{\label{fig:dec_7d}Decoding circuit of the $[[7,1,4]]_{7}$ QRS code
when (a) one qudit, (b) two qudits and (c) three qudits are lost,
respectively. The dashed lines refer to the loss of a qudit. The symbol
``+'' refers to the sum mod 7 gate.}
\end{figure}
 Here, we derive the general approach we use to calculate the fidelity
of the $[[3,1,2]]_{3}$ QRS code when the density matrix of the state
undergoes a dephasing channel. We then derive the terms that depend
on the dephasing for the $[[5,1,3]]_{5}$ and $[[7,1,4]]_{7}$ QRS
codes.

Alice initially encodes her physical state $\left.|\psi\right\rangle _{A}=\alpha_{0}\left.|0\right\rangle +\alpha_{1}\left.|1\right\rangle +\alpha_{2}\left.|2\right\rangle $
into the logical state $\rho_{L}=\left.|\psi\right\rangle _{L}\left\langle \psi|_{L}\right.$
and sends it to Bob over lossy channels, having transmission probability
$p_{1}$ and $p_{2},$ respectively. After the channel loss, the mixed
density matrix can be expressed as a sum of density matrices multiplied
by the loss probability, i.e.,:

\begin{equation}
\begin{array}{c}
\begin{aligned}\rho_{L}\rightarrow\rho' & =p_{1}^{2}p_{2}\rho_{0}+p_{1}p_{2}(1-p_{1})\rho_{1}+p_{1}p_{2}(1-p_{1})\rho{}_{2}\\
 & +p_{1}^{2}(1-p_{2})\rho_{3}+(1-P_{s})I_{3}
\end{aligned}
\end{array}\label{eq:3qudits_loss}
\end{equation}
where $\rho_{0},$ $\rho_{1},$ $\rho_{2}$ and $\rho_{3}$ are the
density matrices resulting from the loss of no qudit, the first qudit,
the second or the third qudit, respectively, while $I_{3}$ is the
normalized identity operator of the Hilbert space spanned by the three
qudits. We assume that the terms of the density matrix $\rho^{'}$
corresponding to the loss of two and three qudits are given by $I_{3}.$
The fidelity of the state given by Eq. \eqref{eq:3qudits_loss} is
therefore a lower bound of the total fidelity. Now, the density matrices,
$\rho_{1}$ and $\rho_{2},$ undergo to a depolarization channel given
by $\rho_{1,2}\rightarrow\rho'_{1,2}=(1-p_{d})\rho_{1,2}+p_{d}I_{1,2}Tr_{12}\left(\rho_{1,2}\right),$
where $I_{1,2}$ is the identity of the Hilbert space spanned by qudit
1 and 2, respectively. Substituting $\rho'_{1,2}$ in Eq. \eqref{eq:3qudits_loss}
we obtain:

\begin{equation}
\begin{array}{c}
\begin{aligned}\rho'\rightarrow\rho(p_{d}) & =p_{1}^{2}p_{2}\rho_{0}+p_{1}p_{2}(1-p_{1})\rho'_{1}\\
 & +p_{1}p_{2}(1-p_{1})\rho'_{2}+p_{1}^{2}(1-p_{2})\rho_{3}+(1-P_{s})I_{3}.
\end{aligned}
\end{array}\label{eq:dep_eq2}
\end{equation}
Bob applies the decoding procedure described in \cite{decoding_LJ},
which ideally restore the initial state of Alice $\left.|\psi\right\rangle _{A}.$
The fidelity, $F,$ is given by $F=_{A}\left\langle \psi|\rho(p_{d})\left.|\psi\right\rangle _{A}.\right.$ 

We now sh ow the derivation of the term $f_{1}(p_{d})$ of Eq. \vref{eq:Fid5_L1mL2}.
To this end, let us assume that the encoded state of the $[[5,1,3]]_{5}$
QRS code losses the first two qudits. The resulting state is then
given by $\rho_{l}=\sum_{i=0}^{5}\left.|\psi\right\rangle _{i}\left\langle \psi|_{i},\right.$
where

$\hspace{-0.5cm}\begin{aligned}\left.|\psi\right\rangle _{0} & \text{\ensuremath{\left.\,=\alpha_{0}|000\right\rangle }+\ensuremath{\left.\alpha_{1}|341\right\rangle }+\ensuremath{\left.\alpha_{2}|132\right\rangle }+\ensuremath{\left.\alpha_{3}|423\right\rangle }}+\left.\alpha_{4}|214\right\rangle ,\\
\left.|\psi\right\rangle _{1} & =\left.\alpha_{0}|111\right\rangle +\left.\alpha_{1}|402\right\rangle +\left.\alpha_{2}|243\right\rangle +\left.\alpha_{3}|034\right\rangle +\left.\alpha_{4}|320\right\rangle \hspace{-0.1cm},\\
\left.|\psi\right\rangle _{2} & =\left.\alpha_{0}|222\right\rangle +\left.\alpha_{1}|013\right\rangle +\left.\alpha_{2}|304\right\rangle +\left.\alpha_{3}|140\right\rangle +\left.\alpha_{4}|431\right\rangle \hspace{-0.1cm},\\
\left.|\psi\right\rangle _{3} & =\left.\alpha_{0}|333\right\rangle +\left.\alpha_{1}|124\right\rangle +\left.\alpha_{2}|410\right\rangle +\left.\alpha_{3}|201\right\rangle +\left.\alpha_{4}|042\right\rangle \hspace{-0.1cm},\\
\left.|\psi\right\rangle _{4} & =\left.\alpha_{0}|444\right\rangle +\left.\alpha_{1}|230\right\rangle +\left.\alpha_{2}|021\right\rangle +\left.\alpha_{3}|312\right\rangle +\left.\alpha_{4}|103\right\rangle \hspace{-0.1cm}.
\end{aligned}
$

We now apply a depolarization channel to $\rho_{l}$ and we obtain
$\rho_{l}\rightarrow\rho'_{l}=(1-p_{d})^{2}\rho_{l}+p_{d}(1-p_{d})I_{1}Tr_{1}(\rho_{l})+p_{d}(1-p_{d})I_{2}Tr_{2}(\rho_{l})+p_{d}^{2}I_{12}Tr_{12}(\rho_{l})$,
where $I_{1(2)}$ is the identity operator of the Hilbert space spanned
by the qudit 1(2) and $I_{12}=I_{1}\otimes I_{2}.$ Bob will apply
the decoding procedure of Fig. \ref{fig:dec_5d} to the state $\rho'_{l}$
obtaining a decoded state $\rho''_{l}$. The contribution of $\rho''_{l}$
to the fidelity of Eq. \eqref{eq:Fid5_L1mL2}, $f_{1}(p_{d})$ will
be:

\begin{equation}
\begin{array}{c}
\begin{aligned}f_{1}(p_{d}) & =_{A}\left\langle \psi|\rho''_{l}\left.|\psi\right\rangle _{A}=\right.\left\langle \psi|_{A}\rho(p_{d})\left.|\psi\right\rangle _{A}\right.\\
 & =\frac{1}{20(\frac{5}{4}-2p_{d}+p_{d}^{2})}\left[4\left(4\sum_{i=0}^{4}\alpha_{i}^{4}+13\sum_{i,j=0}^{4}\alpha_{i}^{2}\alpha_{j}^{2}\right)p_{d}^{2}\right.\\
 & \left.-20\left(2\sum_{i=0}^{4}\alpha_{i}^{4}+5\sum_{i,j=0}^{4}\alpha_{i}^{2}\alpha_{j}^{2}\right)p_{d}+25\right].
\end{aligned}
\end{array}
\end{equation}
It is straightforward to see that $f_{1}(p_{d})$ has a minimum at
$\alpha_{0}=\alpha_{1}=\alpha_{2}=\alpha_{3}=\alpha_{4}=1/\sqrt{5}$
, which is the numerical value we have used to find the fidelities
of Eq. \eqref{eq:Fid5_L1mL2} and \eqref{eq:Fid5L2mL1}. The derivation
of all the other terms due to dephasing follow a similar approach,
hence, we only give here the final result:

$f_{2}(p_{d})=\frac{1}{5-4p_{d}}\left[5-2\left(2\sum_{i=0}^{4}\alpha_{i}^{4}+5\sum_{i,j=0}^{4}\alpha_{i}^{2}\alpha_{j}^{2}\right)p_{d}\right],$

$\begin{array}{c}
\begin{aligned}f_{3}(p_{d}) & =\frac{1}{(5-4p_{d})^{2}}\left[2\left(8\sum_{i=0}^{4}\alpha_{i}^{4}+25\sum_{i,j=0}^{4}\alpha_{i}^{2}\alpha_{j}^{2}\right)p_{d}^{2}\right.\\
 & \left.-20\left(2\sum_{i=0}^{4}\alpha_{i}^{4}+5\sum_{i,j=0}^{4}\alpha_{i}^{2}\alpha_{j}^{2}\right)p_{d}+25\right]
\end{aligned}
\end{array}.$ The contributions to the fidelity that depend on $p_{d}$ of the
$[[7,1,4]]_{7}$ QRS code are:

$\begin{array}{c}
\begin{aligned}g_{1}(p_{d}) & =\frac{342(\frac{7}{6}-p_{d})}{7(49-30p_{d}^{3}+108p_{d}^{2}-126p_{d})}\left[\frac{2}{19}\left(6\sum_{i=0}^{4}\alpha_{i}^{4}\right.\right.\\
 & \left.+19\sum_{i,j=0}^{4}\alpha_{i}^{2}\alpha_{j}^{2}\right)p_{d}^{2}-42\left(2\sum_{i=0}^{4}\alpha_{i}^{4}\right.\\
 & \left.+5\sum_{i,j=0}^{4}\alpha_{i}^{2}\alpha_{j}^{2}p_{d}+\frac{49}{57}\right],
\end{aligned}
\end{array}$

$g_{2}(p_{d})=\frac{1}{6\left(\frac{7}{6}-p_{d}\right)}\left[7-2\left(3\sum_{i=0}^{4}\alpha_{i}^{4}+7\sum_{i,j=0}^{4}\alpha_{i}^{2}\alpha_{j}^{2}\right)p_{d}\right],$

$\begin{array}{c}
\begin{aligned}g_{3}(p_{d}) & =\frac{1}{(6p_{d}-7)^{3}}\left[2\left(108\sum_{i=0}^{4}\alpha_{i}^{4}+343\sum_{i,j=0}^{4}\alpha_{i}^{2}\alpha_{j}^{2}\right)p_{d}^{3}\right.\\
 & -42\left(18\sum_{i=0}^{4}\alpha_{i}^{4}+49\sum_{i,j=0}^{4}\alpha_{i}^{2}\alpha_{j}^{2}\right)p_{d}^{2}\\
 & \left.+294\left(3\sum_{i=0}^{4}\alpha_{i}^{4}+7\sum_{i,j=0}^{4}\alpha_{i}^{2}\alpha_{j}^{2}\right)p_{d}-343\right],
\end{aligned}
\end{array}$

$\begin{aligned}g_{4}(p_{d}) & =\frac{1}{(49-35p_{d}+49p_{d}^{2})}\left[p_{d}^{2}-7\left(5\sum_{i=0}^{4}\alpha_{i}^{4}\right.\right.\\
 & \left.\left.+14\sum_{i,j=0}^{4}\alpha_{i}^{2}\alpha_{j}^{2}\right)p_{d}+49\right],
\end{aligned}
$

$\begin{array}{c}
\begin{aligned}g_{5}(p_{d}) & =\frac{1}{(6p_{d}-7)^{2}}\left[3\left(18\sum_{i=0}^{4}\alpha_{i}^{4}+49\sum_{i,j=0}^{4}\alpha_{i}^{2}\alpha_{j}^{2}\right)p_{d}^{2}\right.\\
 & \left.-28\left(3\sum_{i=0}^{4}\alpha_{i}^{4}+7\sum_{i,j=0}^{4}\alpha_{i}^{2}\alpha_{j}^{2}\right)p_{d}+49\right].
\end{aligned}
\end{array}$

\section{Fidelities of the $[[7,1,4]]_{7},$ QRS code}

Here we give the analytical expressions of the fidelities of the $[[7,1,4]]_{7},$
QRS code case and shows in Fig. \ref{fig:dec_7d}. We label $F_{1(2)}$
the fidelity of the state decoded by Bob when a larger (smaller) number
of qudits is traveling across path 1 (2). 

\subsection{The 6+1 and 1+6 configurations}

\begin{equation}
\begin{array}{c}
\begin{aligned}F_{6+1} & =p_{1}^{6}p_{2}+6p_{1}^{5}p_{2}(1-p_{1})+p_{1}^{6}(1-p_{2})\\
 & +15p_{1}^{4}p_{2}(1-p_{1})^{2}+6p_{1}^{5}(1-p_{1})(1-p_{2})\\
 & +20p_{1}^{3}p_{2}(1-p_{1})^{3}g_{1}(p_{d})\\
 & +15p_{1}^{4}(1-p_{1})^{2}(1-p_{2})+(1-P_{s_{1}})/7^{7}
\end{aligned}
\end{array}\label{eq:6+1}
\end{equation}

and 

\begin{equation}
\begin{array}{c}
\begin{aligned}F_{1+6} & =p_{2}^{6}p_{1}+6p_{2}^{5}p_{1}(1-p_{2})+p_{2}^{6}(1-p_{1})\\
 & +15p_{2}^{4}p_{1}(1-p_{2})^{2}+6p_{2}^{5}(1-p_{2})(1-p_{1})\\
 & +20p_{2}^{3}p_{1}(1-p_{2})^{3}g_{2}(p_{d})\\
 & +15p_{2}^{4}(1-p_{2})^{2}(1-p_{1})+(1-P_{s_{2}})/7^{7},
\end{aligned}
\end{array}\label{eq:1+6}
\end{equation}
where $P_{s_{1(2)}}=p_{1(2)}^{6}p_{2(1)}+6p_{1(2)}^{5}p_{2(1)}(1-p_{1(2)})+p_{1(2)}^{6}(1-p_{2(1)})+15p_{1(2)}^{4}p_{2(1)}(1-p_{1(2)})^{2}+6p_{1(2)}^{5}(1-p_{1})(1-p_{2})+20p_{1(2)}^{3}p_{2(1)}(1-p_{1(2)})^{3}+15p_{1(2)}^{4}(1-p_{1(2)})^{2}(1-p_{2(1)}).$

\subsection{The 5+2 and 2+5 configurations}

\begin{equation}
\begin{array}{c}
\begin{aligned}F_{5+2} & =p_{1}^{5}p_{2}^{2}+5p_{1}^{4}p_{2}^{2}(1-p_{1})+2p_{1}^{5}p_{2}(1-p_{2})\\
 & +10p_{1}^{3}p_{2}^{2}(1-p_{1})^{2}g_{3}(p_{d})+p_{1}^{5}(1-p_{2})^{2}\\
 & +10p_{1}^{4}p_{2}(1-p_{1})(1-p_{2})\\
 & +10p_{1}^{2}p_{2}^{2}(1-p_{1})^{3}g_{4}(p_{d})\\
 & +10p_{1}^{3}p_{2}(1-p_{1})^{2}(1-p_{2})g_{1}(p_{d})\\
 & +5p_{1}^{4}(1-p_{1})(1-p_{2})^{2}+(1-P_{s_{1}})/7^{7}
\end{aligned}
\end{array}\label{eq:5+2}
\end{equation}

and 

\begin{equation}
\begin{array}{c}
\begin{aligned}F_{2+5} & =p_{2}^{5}p_{1}^{2}+5p_{2}^{4}p_{1}^{2}(1-p_{2})+2p_{2}^{5}p_{1}(1-p_{1})\\
 & +10p_{2}^{3}p_{1}^{2}(1-p_{2})^{2}g_{5}(p_{d})+p_{2}^{5}(1-p_{1})^{2}\\
 & +10p_{2}^{4}p_{1}(1-p_{2})(1-p_{1})\\
 & +10p_{2}^{2}p_{1}^{2}(1-p_{2})^{3}g_{4}(p_{d})\\
 & +10p_{2}^{3}p_{1}(1-p_{2})^{2}(1-p_{1})g_{2}(p_{d})\\
 & +5p_{2}^{4}(1-p_{2})(1-p_{1})^{2}+(1-P_{s_{2}})/7^{7},
\end{aligned}
\end{array}\label{eq:2+5}
\end{equation}
where $P_{s_{1(2)}}=p_{1(2)}^{5}p_{2(1)}^{2}+5p_{1(2)}^{4}p_{2(1)}^{2}(1-p_{1(2)})+2p_{1(2)}^{5}p_{2(1)}(1-p_{2(1)})+10p_{1(2)}^{3}p_{2(1)}^{2}(1-p_{1(2)})^{2}+p_{1(2)}^{5}(1-p_{2(1)})^{2}+10p_{1(2)}^{4}p_{2(1)}(1-p_{1})(1-p_{2})+10p_{1}^{2}p_{2}^{2}(1-p_{1(2)})^{3}+10p_{1(2)}^{3}p_{2(1)}(1-p_{1(2)})^{2}(1-p_{2(1)})+5p_{1(2)}^{4}(1-p_{1(2)})(1-p_{2(1)})^{2}.$

\subsection{The 4+3 and 3+4 configurations}

\begin{equation}
\begin{array}{c}
\begin{aligned}F_{4+3} & =p_{1}^{4}p_{2}^{3}+4p_{1}^{3}p_{2}^{3}(1-p_{1})g_{3}(p_{d})\\
 & +6p_{1}^{2}p_{2}^{3}(1-p_{1})^{2}g_{5}(p_{d})+3p_{1}^{4}p_{2}(1-p_{2})^{2}\\
 & +12p_{1}^{3}p_{2}^{2}(1-p_{1})(1-p_{2})g_{3}(p_{d})\\
 & +4p_{1}p_{2}^{3}(1-p_{1})^{3}g_{2}(p_{d})+p_{1}^{4}(1-p_{2})^{3}\\
 & +18p_{1}^{2}p_{2}^{2}(1-p_{1})^{2}(1-p_{2})g_{4}(p_{d})\\
 & +12p_{1}^{3}p_{2}(1-p_{1})(1-p_{2})^{2}g_{1}(p_{d})\\
 & +3p_{1}^{4}p_{2}^{2}(1-p_{2})+(1-P_{s_{1}})/7^{7}
\end{aligned}
\end{array}\label{eq:4+3}
\end{equation}

and

\begin{equation}
\begin{array}{c}
\begin{aligned}F_{3+4} & =p_{2}^{4}p_{1}^{3}+4p_{1}^{3}p_{2}^{3}(1-p_{2})g_{3}(p_{d})\\
 & +6p_{2}^{2}p_{1}^{3}(1-p_{2})^{2}g_{3}(p_{d})+3p_{2}^{4}p_{1}(1-p_{1})^{2}\\
 & +12p_{2}^{3}p_{1}^{2}(1-p_{2})(1-p_{1})g_{5}(p_{d})\\
 & +4p_{2}p_{1}^{3}(1-p_{2})^{3}g_{1}(p_{d})+p_{2}^{4}(1-p_{1})^{3}\\
 & +18p_{2}^{2}p_{1}^{2}(1-p_{2})^{2}(1-p_{1})g_{4}(p_{d})\\
 & +12p_{2}^{3}p_{1}(1-p_{2})(1-p_{1})^{2}g_{2}(p_{d})\\
 & +3p_{2}^{4}p_{1}^{2}(1-p_{1})+(1-P_{s_{2}})/7^{7},
\end{aligned}
\end{array}\label{eq:3+4}
\end{equation}
where $P_{s_{1(2)}}=p_{1(2)}^{4}p_{2(1)}^{3}+4p_{1}^{3}p_{2}^{3}(1-p_{1(2)})+6p_{1(2)}^{2}p_{2(1)}^{3}(1-p_{1(2)})^{2}+3p_{1(2)}^{4}p_{2(1)}(1-p_{2(1)})^{2}+12p_{1(2)}^{3}p_{2(1)}^{2}(1-p_{1})(1-p_{2})+4p_{(2)1}p_{2(1)}^{3}(1-p_{1(2)})^{3}+p_{1(2)}^{4}(1-p_{2(1)})^{3}+18p_{1}^{2}p_{2}^{2}(1-p_{1(2)})^{2}(1-p_{2(1)})+12p_{1(2)}^{3}p_{2(1)}(1-p_{1(2)})(1-p_{2(1)})^{2}+3p_{1(2)}^{4}p_{2(1)}^{2}(1-p_{2(1)}).$ 
\end{document}